\documentstyle[aps]{revtex} \draft \begin{document}
\title{STOCHASTIC RESONANCE IN NONPOTENTIAL SYSTEMS} \author{T.  Alarc\'on,
A.  P\'erez-Madrid, J.M.  Rub\'{\i}} \address{Departament de F\'{\i}sica
Fonamental\\ Facultat de F\'{\i}sica\\ Universitat de Barcelona\\ Diagonal
647, 08028 Barcelona, Spain\\ }

\maketitle

\begin{abstract}

We propose a method to analytically show the possibility for the appearance
of a maximum in the signal-to-noise ratio in nonpotential systems.  We
apply our results to the FitzHugh-Nagumo model under a periodic external
forcing, showing that the model exhibits stochastic resonance.  The
procedure that we follow is based on the reduction to a one-dimensional
dynamics in the adiabatic limit, and in the topology of the phase space of
the systems under study.  Its application to other nonpotential systems is
also discussed.

\end{abstract}


\section{Introduction}

Stochastic resonance ( SR ) \cite{kn:benzi}-\cite{kn:vilar} is a phenomenon
in which an enhancement of the response of a non-linear system is observed
when this system is yielded to an external forcing at some optimized
nonzero noise level.  Since the original proposition of SR as a possible
explanation for periodic recurrences in the global climate dynamics, SR has
become the object of copious theoretical and experimental research in a
wide variety of systems in physics, biology and chemistry.  In all these
works the possibility of noise having beneficial effects in the dynamics of
nonlinear systems has been pointed out.  The original formulation of the
problem, in terms of a bistable system and a periodic forcing has been
extended to systems under the action of aperiodic forcing \cite{kn:collins}
and nondynamical systems \cite{kn:nodin}, \cite{kn:moss2}.

In the present work we focus our attention on nonpotential systems.
Non-potential systems correspond to systems far from equilibrium for which
the principle of detailed balance does not hold.  There are abundant
examples of such systems arising from biological, chemical and physical
problems.  Our contribution in this paper is to develop a formalism which
allows us to analytically treat a wide class of nonpotential systems among
which one can include excitable and threshold systems as well as, for
example, symmetric double-well models \cite{kn:maier}.  In particular, we
apply our approach to continue the work undertaken by some authors in
studying the stochastic properties of the FHN model.  This is a well known
model with wide application in the field of neuronal research
\cite{kn:longtin}, \cite{kn:proceedings}.  Apart from several numerical
simulations done on this subject, Collins $et\;al.$ \cite{kn:collins} have
carried out some analytical work on this matter in the area of aperiodic
stochastic resonance .  Some experimental research has been performed to
show the existence of SR in this model \cite{kn:moss3}.  The results
obtained were compared to the predictions of the theory of SR in
nondynamical systems\cite{kn:moss2}.  Our scheme allows to analytically
approach this problem in a simple way by using a generalization of the
kinetic equations approach used in the case of potential systems ( see
\cite{kn:wiesenfeld}, \cite{kn:reacciones}, \cite{kn:agustin}).

All of the aforementioned models have a common feature; their dynamics
exhibit three fixed points:  an unstable point between two stable points.
This feature established some resemblance between the processes described
by these models and the hopping through a potential barrier.  There are a
great variety of systems that contains these features. The FitzHugh-Nagumo model, in its bistable regime, belongs to them. It is 
worth pointing out that this is not the regime in which this model is used 
to model neural activity. In this context the FHN model is taken to be in the
excitable regime where only one globally attractor exists. As it is pointed 
out by Wiesenfeld et al. \cite{kn:circulo}, a simple model of excitable system consists, among 
other things, of a threshold or potential barrier. Our
theory provides a way to compute the escape rates from the attractors of a 
type of two-dimensional non-potential systems, and therefore it furnishes us 
with the main ingredient to apply the theory developed by Wiesenfeld et al.. 
In this fact you can find the relevance of our work to the field of excitable    systems. Another example which fit the characteristics we are asking for is the
class of symmetric double well systems \cite{kn:maier}.  The Sel'kov model
\cite{kn:ross} for autocatalytic systems gathers these features, too.

This paper is organized as follows.  In Section II we precisely define the
range of applicability of our theory.  We study the dynamics of the
fluctuations and compute the kinetic equations.  In Section III we
introduce the FitzHugh-Nagumo model.  We analyze its stochastic properties
and show the existence of stochastic resonance.  Finally in Section IV we
discuss our main results.

\section{Dynamics of the Fluctuations}

In this work we study the stochastic properties of a class of
two-dimensional nonpotential noisy dynamical systems.  These system may be
characterized by the peculiar topology of the phase space of their
corresponding deterministic underlying versions.  For the case we are
concerned the dynamics is characterized by the presence of three aligned
fixed points; an unstable point between two stable points.  An example of
this kind was given by Maier and Stein \cite{kn:maier} in the context of
the escape problem.  They studied a system with a symmetric phase space
consisting of an hyperbolic point between two sinks whose attraction basins
were separate by the separatrix of the hyperbolic point.

To begin with consider a general two-dimensional noisy dynamical system
\cite{kn:maier2}

\begin{equation}\label{eq:a1}
\frac{d\vec{x}}{dt}\,=\,\vec{u}(\vec{x})\,+\,\vec{\vec{g}}\cdot\vec{\xi}(t),
\end{equation}

\noindent where $\vec{x}$ is the vector whose components are state
variables, the field $\vec{u}(\vec{x})$ is the drift, $\vec{\vec{g}}$ is
the noise matrix and $\xi(t)$ is a gaussian white noise of zero mean and
correlation function given by

\begin{equation}\label{eq:a2} \langle
\xi_{i}(t)\xi_{j}(t')\rangle\,=\,2D\delta_{ij}\delta(t-t'), \end{equation}

\noindent with $D$ being the noise intensity.  For the case discussed in
\cite{kn:maier} the components of the drift are

\begin{equation}\label{eq:a3} u_{1}(\vec{x})\,=\,P_{3}(x)+a{x}^{m}y+b,
\end{equation}

\begin{equation}\label{eq:a4} u_{2}(\vec{x})\,=\,cx-dy+e, \end{equation}

\noindent where $P_{3}(x)$ is a third order polynomial, and $m$ is an
integer such that $0\leq m\leq 2$.  From the second of these equations, it
is easy to check that all the fixed points are aligned.  By equating eq.
(\ref{eq:a3}) to zero, one obtains a third order equation with three real
roots for the proper values of the parameters.

The corresponding Fokker-Planck equation for the probability density,
$\rho(x,t)$, is

\begin{equation}\label{eq:a5} \frac{\partial\rho}{\partial
t}\,=\,\nabla\cdot(-\vec{u}(\vec{x})\rho+\nabla\cdot(\vec{\vec{D}}\rho)),
\end{equation}

\noindent where $\vec{\vec{D}}\,=\,D\vec{\vec{g}}\cdot\vec{\vec{g}}^{T}$ is
the diffusion tensor.

Let us now assume that the system is potential.  In such a case it is
possible to write the Fokker-Planck equation as a continuity equation:

\begin{equation}\label{eq:a6} \frac{\partial\rho}{\partial
t}\,=\,-\nabla\cdot\vec{J}, \end{equation}

\noindent where $\vec{J}$ is the diffusion current given by

\begin{equation}\label{eq:a7} \vec{J}\,=\,-D\,e^{-U/D}\nabla\cdot
e^{\vec{\vec{\mu}}/D}.  \end{equation}

\noindent To obtain this expression, we have defined $U$ and
$\vec{\vec{\mu}}$ ( a generalized chemical potential) as follows

\begin{equation}\label{eq:a8} U\,\equiv\,-\int\,\vec{u}\cdot d\vec{x},
\end{equation}

\begin{equation}\label{eq:a9}
e^{\vec{\vec{\mu}}/D}\,\equiv\,\vec{\vec{g}}\rho e^{U/D}.  \end{equation}

For a potential system, the function $U$ as defined in (\ref{eq:a8}) is
simply the potential energy.  In the nonpotential case, however, the value
of $U$ will depend on the path of integration we choose and, in general, we
can not achieve eqs.  (\ref{eq:a6}) and (\ref{eq:a7}).

Now consider the weak noise limit and think about some general
characteristics of the probability distribution.  If we let the system
evolve during a sufficiently long time the probability distribution will
have two maxima at the two stable fixed points ( SFP ) of the deterministic
dynamics and a minimum in the unstable fixed point ( USFP )
\cite{kn:ebeling}.  On the other hand, in the weak noise limit the
probability distribution will be very narrow around the line on which the
fixed point lies.  Thus we can assume that the fluctuations run over this
line and, therefore, their dynamics are practically one-dimensional.
This approximation can be justified accounting for the assumption of low 
noise level and the adiabatic limit. As well as in the one-dimensional double-well problem the adiabatic hypothesis implies that the representative point of the system is, in this long time limit, in one of the two wells \cite{kn:wiesenfeld}, \cite{kn:reacciones}, in our two dimensional problem implies that the dynamics will be restricted to be on the nullcline. On the other hand, as it was found by Maier and Stein in \cite{kn:maier}, the distance to the nullcline, in the limit of weak noise intensity, is normally distributed with variance equal to the square root of noise level. Therefore, the fluctuations will run in a very narrow region around the nullcline.
Consequently, although the whole system is nonpotential, we can reduce the
dynamics being one-dimensional whose potential is given by the function $U$
taking as the integration path the null cline.  For the case discussed
previously from eqs.  (\ref{eq:a3}), (\ref{eq:a4}) and (\ref{eq:a8}), under
these circumstances ,one then obtains the drift

\begin{equation}\label{eq:a10}
u_{1}(x)\,=\,P_{3}(x)+\frac{ac}{d}{x}^{m+1}+b+\frac{eax^m}{d}.  \end{equation}

\noindent corresponding to the dynamics of the fluctuations in the weak
noise limit and

\begin{equation}\label{eq:a11}
U\,=\,-\int^{x}\,dt(\,P_{3}(t)+\frac{ac}{d}{t}^{m+1}+b+\frac{eax^m}{d}),
\end{equation}

Our next step will be to discretize the dynamics on the null cline .  In
particular we will obtain the kinetic equations.  To this end we define the
populations $n_{+}$ ( $n_{-}$ ) as the population on the right ( left ) of
the USFP \cite{kn:gardiner}.

\begin{equation}\label{eq:a12}
n_{+}\,=\,\int_{\it{S[+]}}\,\rho(\vec{x},t)d\vec{x} \end{equation}

\begin{equation}\label{eq:b2}
n_{-}\,=\,\int_{\it{S[-]}}\,\rho(\vec{x},t)d\vec{x} \end{equation}

\noindent where $\it{S[+]}$ ( $\it{S[-]}$ ) is the portion of the phase
space on the right( left ) of a line orthogonal to the line which contains
the fixed points and passing through the USFP .

In the adiabatic limit we can assume that the population is strongly
concentrated in a small region around the SFP, as suggested by the picture
of the probability density that we have profiled when the maxima in this
long time limit is very high.  This corresponds, in the assumption of
one-dimensional fluctuations dynamics, having a bistable potential with two
deep minima at the SFP and maximum at the USFP, or equivalently a high
potential barrier.

In the present context we understand by adiabatic aproximation a long time 
limit in which all the system has arrived to a quasistationary state 
such that the probability of the system to be in a state different from the 
stationary stable states is practically zero.

So in this limit we can suppose that the system reaches a quasi-stationary
state in which a quasi-stationary diffusion current is established.  In
addition, is assumed to be uniform between the two maxima of the
probability density and, in the weak noise limit this current is
concentrated in the line joining the three fixed points without loss of
generality, the system can be taken to lie in the axis $y=0$.

\begin{equation}\label{eq:b3}
\vec{J}(\vec{x},t)\,=\,\vec{J}(x,t)\delta(y)\,=\,\vec{J}(t)
\delta(y)(\theta(x-x_{+})-\theta(x-x_{-})), \end{equation}

\noindent where $x_{+}$ ( $x_{-}$ ) is the coordinate of the fixed point on
the right ( left ) of the USFP.

The kinetic equation for $n_{+}$ is given by

\begin{equation}\label{eq:b4}
\frac{dn_{+}}{dt}\,=\,\int_{\it{S[+]}}\,\frac{\partial\rho}{\partial
t}d\vec{x}\,=\,-\int_{\it{S[+]}}\,\nabla\cdot \vec{J} d\vec{x}.
\end{equation}

\noindent By using the divergence theorem and the assumptions about the
form of the diffusion current, we have

\begin{equation}\label{eq:b5}
\frac{dn_{+}}{dt}\,=\,\int_{-\infty}^{+\infty}J_{1}(x,t).
\delta(y)dy\,=\,J_{1}(x,t), \end{equation}

\noindent and, proceeding in the same way for $n_{-}$, we obtain

\begin{equation}\label{eq:b6} \frac{dn_{-}}{dt}\,=\,-J_{1}(x,t)
\end{equation}

In addition, due to the height of the maxima in the probability density and
the weakness of the noise, we can also consider that equilibrium in each
side of the USFP is reached independently, thus, the generalized chemical
potential is given by

\begin{equation}\label{eq:b7}
\mu(\vec{x},t)\,=\,\{\mu(x_{+},t)\theta(x_{0}-x)+
\mu(x_{-},t)\theta(x-x_{0})\}\delta(y), \end{equation}

\noindent where $x_{0}$ is the coordinate of the USFP.  The tensorial
character of the generalized potential has been removed due to the dynamics
reduction to one dimension.  In equation (\ref{eq:b7}) $\mu\equiv\mu_{11}$
has been defined.  By using equation (\ref{eq:b7}) in (\ref{eq:a7}) we
obtain

\begin{equation}\label{eq:b8}
\Psi(\vec{x},t)\,=\,\Psi_{+}e^{-(U-U_{+})/D}\theta(x_{0}-
x)+\Psi_{-}e^{-(U- U_{-})/D}\theta(x-x_{0})\}\delta(y), \end{equation}

\noindent where $U$ corresponds to the integral over the adequate path of
(\ref{eq:a10}), $U_{+}$ and $U_{-}$ are its values at the SFP ( its minima
), $\Psi\equiv g_{11}\rho$ and $\Psi_{\pm}=g_{11}\rho(x_{\pm},t)$.

In order to obtain the expression for the quasi-stationary current
$J_{1}(t)$ we follow the same procedure as in \cite{kn:reacciones}.  From
the definition of the probability current and the adiabatic hypothesis we
have

\begin{equation}\label{eq:b9} J_{1}(t)(\theta(x-x_{+})-\theta(x-x_{-}))\,=
\,-De^{-U/D}\partial_{x}e^{\mu/D}, \end{equation}

\noindent where a diagonal diffusion tensor is assumed.  Integrating now
over $x$ and taking into account that, due to the height of the barrier,
the main contribution to these integrals is around the maximum of the
potential $x_{0}$, one obtains

\begin{equation}\label{eq:b10} J_{1}(t)\,=\,-D(\frac{\vert
U_{0}^{\prime\prime}\vert}{2\pi
D})^{1/2}e^{-U_{0}/D}(e^{\mu^{+}/D}-e^{\mu^{-}/D}), \end{equation}

\noindent where $U_{0}^{^{\prime\prime}}\equiv\frac{d^{2}U}{dx^2}
\vert_{x_{0}}$, $U_{0}\equiv U(x_{0})$ and $\mu^{+}$, $\mu^{-}$ are the
values of the generalized chemical potential at SFP.

By using eq.  (\ref{eq:b8}) we can rewrite (\ref{eq:b10}) in the following
way

\begin{equation}\label{eq:b11} J_{1}(t)\,=\,D(\frac{\vert
U_{0}^{^{\prime\prime}}\vert}{2\pi
D})^{1/2}(\Psi_{-}e^{-(U_{0}-U_{-})/D}-\Psi_{+} e^{-(U_{0}-U_{+})/D}),
\end{equation}

\noindent where $U_{+}$ and $U_{-}$ are the values of the potential at the
SFP.  On the other hand, by using eq.  (\ref{eq:b8}) in eq.  (\ref{eq:a12})
one obtains the following relations

\begin{equation}\label{eq:b12}
\Psi_{+}\,=\,(\frac{U_{+}^{^{\prime\prime}}}{2\pi D})^{1/2}n_{+},
\end{equation}

\begin{equation}\label{eq:b13}
\Psi_{-}\,=\,(\frac{U_{-}^{^{\prime\prime}}}{2\pi D})^{1/2}n_{-}.
\end{equation}

These expressions allow us to write the kinetic equations for the
populations $n_{+}$ and $n_{-}$ in the form

\begin{equation}\label{eq:b14}
\frac{dn_+}{dt}=-\frac{dn_-}{dt}\,=\,K_{-}n_{-}-K_{+}n_{+}, \end{equation}

\noindent where the kinetic coefficients are given by

\begin{equation}\label{eq:b15} K_{\mp}\,=\,\frac{1}{2\pi}(\vert
U_{0}^{^{\prime\prime}}\vert U_{\mp}^{^{\prime\prime}})^{1/2}
e^{-(U_{0}-U_{\mp})/D}, \end{equation}

With this result we have obtained of the kinetic equations for the
nonpotential system.

\section{The FitzHugh-Nagumo Model}

In this section, we will apply the results of the previous section to the
study of the FitzHugh-Nagumo ( FHN ) neural model \cite{kn:longtin},
\cite{kn:fitz}-\cite{kn:murray}.  This model is a variant of the
Hodgkin-Huxley model \cite{kn:murray}-\cite{kn:hille} which accounts for
the essentials of the regenerative firing mechanisms in nerve cells.  The
FHN equations correspond to an excitable threshold model but, as will be
seen briefly, due to their cubic non-linearity, they exhibit the
characteristic behavior of a bistable system.  Our main objective is to
show analytically the appearance of SR in this model under a periodic
external forcing.

The non-dimensional equations of the FHN model are \cite{kn:murray}

\begin{equation}\label{eq:c1} \frac{dv}{dt}\,=\,v(a-v)(v-1)-w+I_{a},
\end{equation}

\begin{equation}\label{eq:c2} \frac{dw}{dt}\,=\,bv-\gamma w.
\end{equation}

\noindent where $0<a<1$ is essentially the threshold value, $b$ and
$\gamma$ are positive constants and $I_{a}$ is the applied current.  For
the sake of simplicity, and without loss of generality, we will take
$I_{a}=0$.  The drift field for this model is given by

\begin{equation}\label{eq:c3} u_{1}(v,w)\,=\,v(a-v)(v-1)-w \end{equation}

\begin{equation}\label{eq:c4} u_{2}(v,w)\,=\,bv-\gamma w \end{equation}

As can be seen from eq.  (\ref{eq:c2}) the null cline of the deterministic
dynamics of this equations is the line $v=\frac{\gamma}{b}w$.  By
substitution on the right hand side of equation (\ref{eq:c1}) we find the
following equation for steady states

\begin{equation}\label{eq:c5} v(a-v)(v-1)-\frac{b}{\gamma}v=0.
\end{equation}

\noindent This is a third order equation, which for certain values of the
parameters has three roots ( see Figure 1 ).  Among these three fixed
points two are stable:  $F_{-}$ and $F_{+}$, and one unstable:  $F_{0}$,
situated between the other two \cite{kn:murray}.  Thus, in this case, this
system fulfills the conditions in order to our theory to be applied.

The function $U$ defined in Section II has to be integrated in this case
over line $v=\frac{\gamma}{b}w$.  Performing this integration one obtains

\begin{equation}\label{eq:c6}
U=\frac{v^{4}}{4}-\frac{(a+1)}{3}v^{3}+\frac{(a+b/\gamma)}{2}v^{2}.
\end{equation}

On the other hand, when this system is in a noisy environment, in the limit
of weak noise, we can approximate the dynamics of the fluctuations by the
one-dimensional Langevin equation

\begin{equation}\label{eq:c7}
\frac{dv}{dt}\,=\,v(a-v)(v-1)-\frac{b}{\gamma}v+\xi(t), \end{equation}

\noindent that is, the fluctuations run along the line
$v=\frac{\gamma}{b}w$.  As can be easily checked, $U$ is the potential for
the deterministic part of this equation.  The two SFP of this
one-dimensional dynamics are the two minima of (\ref{eq:c6}) and the USFP
is its maximum.  Collins $\it{et\,\,al.}$ \cite{kn:collins} have arrived
previously to similar conclusions by another approach in the context of the
study of aperiodic stochastic resonance.

Fig.  2  shows the asymmetric shape of $U$.  Before going any further, it
would be interesting to summarize what this picture has to say to us about
the physics of the problem \cite{kn:longtin}, \cite{kn:collins}.  In the
FHN model there is a fast variable, $v(t)$, and a recovery-like variable,
$w(t)$.  After the barrier threshold has been crossed, i.e.  the system has
gone to an "excited" state, the system returns ( even in the deterministic
case ) to the "rest" state.  As can be seen in Fig.  2, there is one of the
stable states for which the potential is larger than for the other stable
state.  Therefore, there is a more stable state, which corresponds to the
rest state to which the system after some time , under the action of the noise, returns. In our scheme the presence of noise is necessary in order to return to the rest state, because of the elimenation of $w(t)$.  

In order to show how this scheme can account for the existence of
stochastic resonance in the FHN model, we will suppose that the system is
under the action of periodic forcing.  For simplicity's sake, we will
assume that the parameter $a$ is a periodic function:
$a=a_{0}(1+\epsilon_{0} sin\omega_{s}t)$ where $\epsilon_{0}$ is a small
parameter and $a_{0}(1+\alpha)<1$.

To take $a$ as an oscillatory factor implies that the positions of $F_{0}$
( the USFP ) and $F_{+}$ ( one of the SFP ) as well as the values of the
potential at these points become oscillatory functions, too.  The position
of $F_{-}$ (the other SFP ) remains constant.  Let $v_{0}$, $v_{-}$ and
$v_{+}$ be the v-coordinate of the maximum and the minima, respectively, of
the potential; one has

\begin{equation}\label{eq:c13}
U_{+}\,\equiv\,U(v_{+})\,=\,\xi_{+}+\eta_{+}\epsilon(t).  \end{equation}

To compute the moments and the power spectrum, we assume that, in the limit
of weak noise, the probability density ( in one dimension ) can be written
as \cite{kn:wiesenfeld}

\begin{equation}\label{eq:c14}
p(v,t)\,=\,n_{+}(t)\delta_{v,v_{+}}+n_{-}(t)\delta_{v,v_{-}},
\end{equation}

\noindent where $n_{+}$ and $n_{-}$ come from the kinetic equations
(\ref{eq:b14}).  The formal solution to these equations is found to be

\begin{equation}\label{eq:c15}
n_{\pm}(t)\,=\,g^{-1}(t)(n_{\pm}(t_{0})g(t_{0})+\int_{t_{0}}^{t}\,
K_{\mp}(t')g(t')dt'), \end{equation}

\noindent with

\begin{equation}\label{eq:c16} g(t)\,=\,\int^{t}(K_{+}+K_{-})dt'.
\end{equation}

In order to calculate $n_{\pm}$ we perform a Taylor expansion of the
transition rates up to the first order in respect to the parameter
$\epsilon(t)$.

\begin{equation}\label{eq:c17}
K_{\pm}\,=\,\alpha_{0}^{\pm}+\alpha_{1}^{\pm}\phi_{0}sin\omega_{0}t
\end{equation}

\noindent where $\phi_{0}$ has been defined as
$\phi_{0}\equiv\epsilon_{0}/D$ and $\alpha_{0}^{\pm}$ and
$\alpha_{1}^{\pm}$ are given by

\begin{equation}\label{eq:c18}
\alpha_{0}^{\pm}\,=\,\frac{e^{-\xi_{0}/D}}{2\pi}d_{0}^{\pm}e^{\xi_{\pm}/D},
\end{equation}

\begin{equation}\label{eq:c19} \alpha_{1}^{\pm}\,=
\,\frac{e^{-\xi_{0}/D}}{2\pi}
d_{0}^{\pm}e^{\xi_{\pm}/D}(\eta_{0}-\eta_{\pm}), \end{equation}

\noindent with $d_{0}^{\pm}$ being the zero order contribution of $\vert
U_{0}^{\prime\prime}\vert U_{\pm}^{\prime\prime}$, $\xi_{0}$ and $\eta_{0}$
the zero and first order contribution of $U$ at the USFP and $\xi_{\pm}$
and $\eta_{\pm}$ the zero and first order contribution of $U$ at the SFPs.
Its first order contribution can be neglected in the limit of weak noise.
By introducing these Taylor expansions in equations (\ref{eq:c15}) and
(\ref{eq:c16}), we get the populations up to the first order in the
parameter $\phi_{0}$.

\begin{eqnarray} n_{\pm}(t)\,
&=&\,e^{-\alpha_{0}(t-t_{0})}(\delta_{v(t_{0}),v_{\pm}}-
\frac{\alpha_{0}^{\mp}}{\alpha_{0}}-\phi_{0}\frac{\alpha_{0}^{\mp}
\alpha_{1}}{\alpha_{0}\omega_{0}}cos(\omega_{0}t)+\phi_{0}
\alpha_{1}^{\mp}\frac{cos(\omega_{0}t_{0}+\Phi)}{(\alpha_{0}^{2}
+\omega_{0}^{2})^{1/2}}-\nonumber\\ &
&\phi_{0}\frac{\alpha_{1}}{\omega_{0}}
\frac{cos(\omega_{0}t_{0}+\Phi)}{(\alpha_{0}^{2}+
\omega_{0}^{2})^{1/2}})+\frac{\alpha_{0}^{\mp}}{\alpha_{0}}
(1+\phi_{0}\frac{\alpha_{1}}{\omega_{0}}cos(\omega_{0}t))-
\phi_{0}\alpha_{1}^{\mp}\frac{cos(\omega_{0}t+\Phi)}{(\alpha_{0}^{2}
+\omega_{0}^{2})^{1/2}}+\nonumber \\ &
&\phi_{0}\frac{\alpha_{1}}{\omega_{0}}
\frac{sin(\omega_{0}t+\Phi)}{(\alpha_{0}^{2}+\omega_{0}^{2})^{1/2}},
\end{eqnarray}

\noindent where $\Phi\equiv arctg(\alpha_{0}/\omega_{0})$ and $\alpha_{0}=\alpha_{0}^{+}+\alpha_{0}^{-}$. The quantity
$n_{\pm}(t\vert v_{t_{0}},t_{0})$ is the conditional probability that
$v(t)$ is in the + state at time $t$ given that the state at time $t_{0}$
was $v_{t_{0}}$.  From this equation it is possible to compute the
statistical properties of the process $v(t)$.  Of particular interest to
our purposes is to find its autocorrelation function which is given by
(Wiesenfeld)

\begin{equation}\label{eq:c20} \langle
v(t)v(t+\tau)\rangle\,=\,lim_{t\rightarrow -\infty}\langle
v(t)v(t+\tau)\vert v_{t_{0}},t_{0} \rangle.  \end{equation}

\noindent The conditional correlation function is given by

\begin{eqnarray} \langle v(t)v(t+\tau)\vert v_{t_{0}},t_{0}\rangle
&=&v_{+}(t+\tau)v_{+}(t)n_{+}(t+\tau\vert v_{+},t)n_{+} (t\vert
v_{t_{0}},t_{0})+\nonumber\\ & &v_{+}(t+\tau)v_{-}(t)n_{+}(t+\tau\vert
v_{-},t)n_{-}(t\vert v_{t_{0}},t_{0})+\nonumber\\ &
&v_{-}(t+\tau)v_{+}(t)n_{-}(t+\tau\vert v_{+},t)n_{+}(t\vert
v_{t_{0}},t_{0})+ \nonumber \\ & &v_{-}(t+\tau)v_{-}(t)n_{-}(t+\tau\vert
v_{+},t)n_{-}(t\vert v_{t_{0}},t_{0}).  \end{eqnarray}

Let us make some considerations which allows us to simplify the computation
of the autocorrelation function.  It is clear from equation (43) that the
Fourier transform of the autocorrelation function will depend on $t$ as
well as in the frequency.  This dependency is avoided by taking its average
over the period of the external forcing \cite{kn:wiesenfeld}.  The
autocorrelation function is to be computed up to the second order in
parameter $\phi_{0}\sim D^{-1}$.Thus, in the limit of the weak noise
$D^{-1}$, can be neglected in comparision to $D^{-2}$.  Therefore, the only
contribution of the first order term of $v_{\pm}$ to the autocorrelation
function comes from its product with the zero order term of the product of
$n$'s.  But, on doing the average this term vanishes.  So, finally, and
taking into account that $v_{-}=0$, we arrive to

\begin{equation}\label{eq:c21} \overline{\langle v(t)v(t+\tau)\vert
v_{t_{0}},t_{0}\rangle}\,=\,\lambda_{+}^{2} \overline{n_{+}(t+\tau\vert
v_{+},t)n_{+}(t\vert v_{t_{0}},t_{0})}, \end{equation}

\noindent where the overlined indicates an average over $t$ and
$\lambda_{+}$ is the position of $F_{+}$ up to order zero.  From equations
(43) and (\ref{eq:c21}) taking the average and the limit $t_{0}\rightarrow
-\infty$, we finally obtain the following expression for the
autocorrelation function.

\begin{eqnarray} \overline{\langle v(t)v(t+\tau)\rangle}\,
&=&\,\lambda_{+}^{2}
(\frac{\alpha_{0}^{-}}{\alpha_{0}})^{2}+e^{-\alpha_{0}\vert\tau\vert}
(\frac{\alpha_{0}^{-}}{\alpha_{0}})(1 -\frac{\alpha_{0}^{-}}{\alpha_{0}})+
\nonumber \\ &
&\lambda_{+}^{2}\phi_{0}^{2}\{\frac{1}{2}e^{-\alpha_{0}\vert\tau\vert}
(\frac{\alpha_{0}^{-}\alpha_{1}^{-}\alpha_{1}}{\alpha_{0}}-\alpha_{0}^{-}
(\frac{\alpha_{1}}{\omega_{0}})^{2}-(\alpha_{1})^{2}-
(\frac{\alpha_{1}}{\omega_{0}})^{2})(\frac{1}{\alpha_{0}^{2}+
\omega_{0}^{2}})+ \nonumber \\ &
&(\frac{\alpha_{0}^{-}\alpha_{1}^{-}\alpha_{1}}{\alpha_{0}}-\alpha_{0}
(\frac{\alpha_{1}}{\omega_{0}})^{2})\frac{cos(\omega_{0}\tau)}{\alpha_{0}
^{2}+\omega_{0}^{2}}+ \nonumber \\ &
&(\frac{\alpha_{0}^{-}}{\alpha_{0}}\frac{\alpha_{1}^{2}}{\omega_{0}}
+\frac{\alpha_{0}^{-}\alpha_{1}^{-}\alpha_{1}}{\alpha_{0}})
sin(\omega_{0}\tau))+ \nonumber \\ &
&\frac{cos(\omega_{0}\tau)}{\alpha_{0}^{2}+\omega_{0}^{2}}
(\frac{\alpha_{0}^{-}\alpha_{1}^{2}}{\omega_{0}^{2}}-
\frac{\alpha_{0}^{-}\alpha_{1}^{-}\alpha_{1}}{\alpha_{0}}+
\frac{1}{2}(\alpha_{1}^{-})^{2}+\frac{1}{2}
(\frac{\alpha_{1}}{\omega_{0}})^{2})\}.  \end{eqnarray}

With this results we can now compute the averaged power spectrum given by

\begin{equation}\label{eq:c23}
\overline{S(\Omega)}\,=\,\frac{\omega_{0}}{2\pi}\int_{0}^{2\pi/\omega_{0}}
S(\Omega ,t)dt\,=\,\int_{-\infty}^{+\infty}\overline{\langle
v(t)v(t+\tau)\rangle}e^{-\imath\Omega\tau}d\tau, \end{equation}

\noindent where the last equality follows from the commutative character of
averaging and Fourier transform.  We obtain, after Fourier transform (50)

\begin{eqnarray} \overline{S(\Omega)}\,
&=&\,\lambda_{+}^{2}\frac{\alpha_{0}^{-}}{\alpha_{0}}
\delta(\Omega)+2\frac{\lambda_{+}^{2}}{\alpha_{0}^{2}+
\Omega^{2}}\alpha_{0}^{-}(1-\frac{\alpha_{0}^{-}}{\alpha_{0}})
+2\pi\frac{\lambda_{+}^{2}\phi_{0}^{2}}{\alpha_{0}^{2}+\Omega^{2}}
(\frac{(\alpha_{0}^{-})^{2}\alpha_{1}^{2}}{\omega_{0}^{2}}- \nonumber \\ &
&\frac{(\alpha_{0}^{-})^{2}\alpha_{1}^{-}\alpha_{1}}{\alpha_{0}^{2}}+
\frac{1}{2}(\frac{\alpha_{1}^{-1})^{2}}{\alpha_{0}^{-}}+\frac{1}{2}
(\frac{\alpha_{1}\alpha_{0}^{-}}{\omega_{0}})^{2})
\delta(\Omega-\omega_{0}).  \end{eqnarray}

\noindent In this expression the fraction of the total power in the
broadband noisy part of the spectrum, which usually is a small fraction of
the total power, has been neglected \cite{kn:wiesenfeld}.  Note that in the
power spectrum there is a term proportional to $\delta(\Omega)$.  This is
due to the asymmetry of the potential that originates a mean probability
current between one stable state and the more stable one is more stable
than the other one.  This term has not been obtained in previous
approaches to this problem \cite{kn:collins}, \cite{kn:moss2}

From equation (47) the signal to noise ratio, R, can be obtained as a
function of the noise level $D$, by making $\Omega=\omega_{0}$.

\begin{equation}\label{eq:c25}
R\,=\,\frac{\frac{\epsilon_{0}^{2}\pi}{D^2}(\frac{3}{2}
\alpha_{0}^{-}(\frac{\alpha_{1}}{\omega_{0}})^{2}-
\frac{\alpha_{0}^{-}\alpha_{1}^{-}\alpha_{1}}{\alpha_{0}^{2}}
+\frac{1}{2}\frac{(\alpha_{1}^{-})^{2}}{\alpha_{0}^{-}})}{1-
\frac{\alpha_{0}^{-}}{\alpha_{0}}}.  \end{equation}

We have plotted R in Figure 3 as a function of the noise level.  It
exhibits a maximum for a non-zero noise level and therefore this model
shows stochastic resonance.

\section{Conclusions}

In this paper we have proposed a method to analyze the possibility for the
appearance of SR in nonpotential systems.  In particular we have treated
the FHN model.  A lot of numerical work on this subject has been performed
but only a few papers have treated this problem from an analytical
perspective due to the difficulties inherent to the nature of nonpotential
system.  The work by Collins $et\;al$ \cite{kn:collins} where aperiodic SR
is discussed, and the paper by Pei $et\;al.$ \cite{kn:moss3}, in which the
theoretical framework developed by Gingl $et\;al.$ \cite{kn:moss2} was used
to interpret some experimental results, are among the ones discussing
analytical aspects.

Far from being specific of this problem, the method we have proposed can be
extended to a wide variety of different nonpotential systems ranging from
threshold systems to some autocatalytic models and symmetric double well
systems from fields so diverse as chemical kinetics and population
dynamics. Although, for the sake of brevity, we do not quote our results here we have applied our approach to that of Collins $et\;al.$ \cite{kn:collins} 
with a periodic input and we obtain similar results in both cases, not the
same because in the case of these authors the input is additive and the 
threshold is maintained constant. We have applied our theory to the standard
double-well model \cite{kn:maier} and we have obtain that this model exhibits
SR. The result in this case is equivalent to the one obtained in 
\cite{kn:wiesenfeld} for a symmetric quartic potential.

The scheme we present in this paper allows us to treat potential and the
class of aforementioned nonpotential systems in an unified way.  Moreover,
it reproduces the essentials of the physics of the problem as can be seen
from the obtaining of the refractory current in eq.  (52)
\cite{kn:longtin}.

Finally, it is worth pointing out that our method, which constitutes
essentially an extension of the Kramers rate theory to this kind of
nonpotential systems, enables one to compute the kinetic coefficients in a
simple and direct way.

\acknowledgments

This work has been supported by DGICYT of the Spanish Government under
grant PB95-0881.  One of us (T.  Alarc\'on) wishes to thank to DGICYT of
the Spanish Government for financial support.

\newpage

\vspace{3cm} \large

\begin{center} FIGURE CAPTIONS \end{center} \begin{itemize}

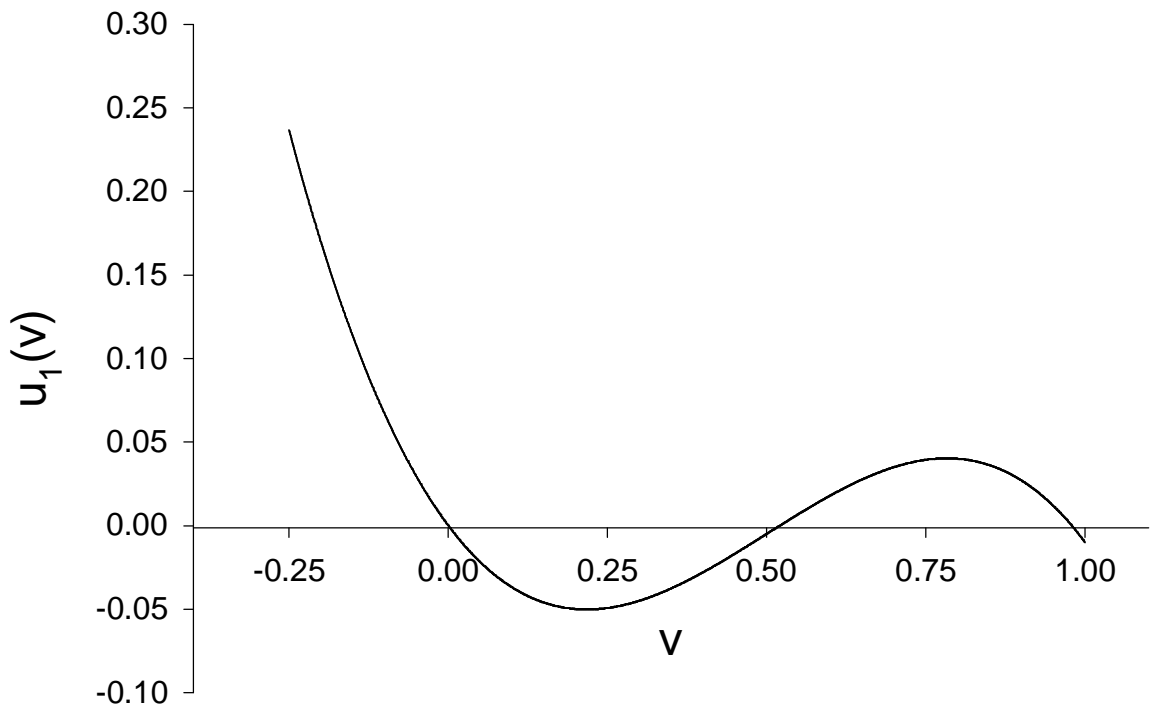
\item{Figure 1}.- Fixed points of the FitzHugh-Nagumo model for $a=0.5$ and
$\frac{b}{\gamma}=0.01$. Non-dimensional variables.

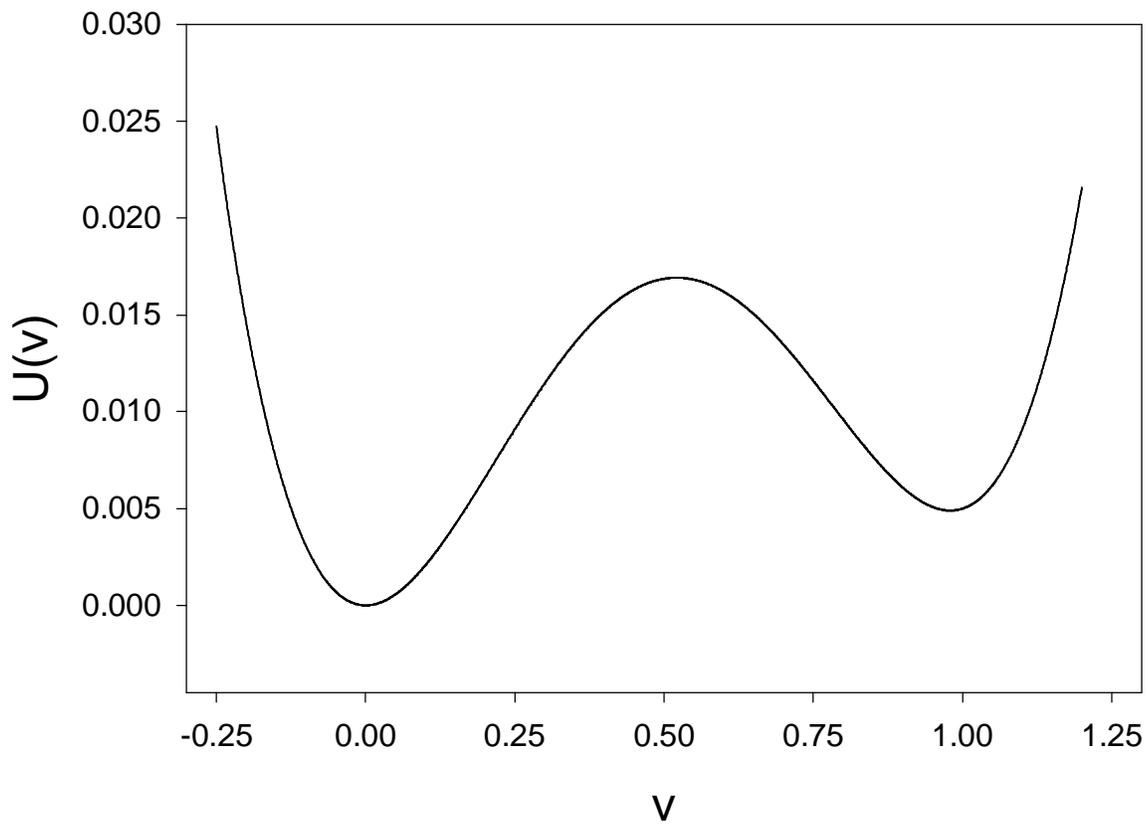
\item{Figure 2}.- Asymmetric one-dimensional potential as a function of
$v$. Non-dimensional variables.

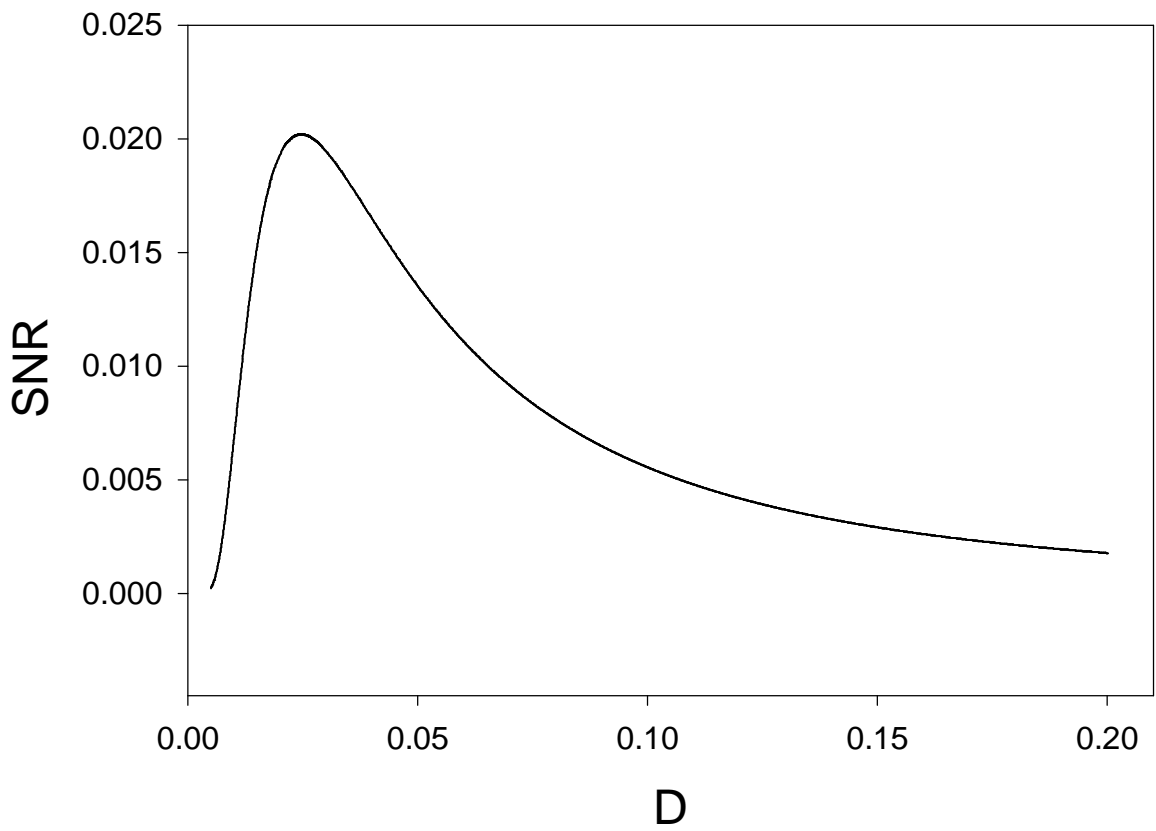
\item{Figure 3}.- Signal-to-Noise Ratio as a function of the noise level,
$D$. Non-dimensional variables.

\end{itemize}

\end{document}